\documentclass[amsmath,amssymb,aps,pra,twocolumn,10pt,superscriptaddress,showpacs]{revtex4-1}

\def\bra#1{\mathinner{\langle{#1}|}}
\def\ket#1{\mathinner{|{#1}\rangle}}
\def\braket#1{\mathinner{\langle{#1}\rangle}}

\let\protect\relax

{\catcode`\|=\active
  \xdef\Braket{\protect\expandafter\noexpand\csname Braket \endcsname}
  \expandafter\gdef\csname Braket \endcsname#1{\begingroup
     \ifx\SavedDoubleVert\relax
       \let\SavedDoubleVert\|\let\|\BraDoubleVert
     \fi
     \mathcode`\|32768\let|\BraVert
     \left\langle{#1}\right\rangle\endgroup}
}
\def\BraVert{\@ifnextchar|{\|\@gobble}
     {\egroup\,\mid@vertical\,\bgroup}}
\def\BraDoubleVert{\egroup\,\mid@dblvertical\,\bgroup}
\let\SavedDoubleVert\relax

{\catcode`\|=\active
  \xdef\set{\protect\expandafter\noexpand\csname set \endcsname}
  \expandafter\gdef\csname set \endcsname#1{\mathinner
        {\lbrace\,{\mathcode`\|32768\let|\midvert #1}\,\rbrace}}
  \xdef\Set{\protect\expandafter\noexpand\csname Set \endcsname}
  \expandafter\gdef\csname Set \endcsname#1{\left\{
     \ifx\SavedDoubleVert\relax \let\SavedDoubleVert\|\fi
     \:{\let\|\SetDoubleVert
     \mathcode`\|32768\let|\SetVert
     #1}\:\right\}}
}
\def\midvert{\egroup\mid\bgroup}
\def\SetVert{\@ifnextchar|{\|\@gobble}
    {\egroup\;\mid@vertical\;\bgroup}}
\def\SetDoubleVert{\egroup\;\mid@dblvertical\;\bgroup}

\begingroup
 \edef\@tempa{\meaning\middle}
 \edef\@tempb{\string\middle}
\expandafter \endgroup \ifx\@tempa\@tempb
 \def\mid@vertical{\middle|}
 \def\mid@dblvertical{\middle\SavedDoubleVert}
\else
 \def\mid@vertical{\mskip1mu\vrule\mskip1mu}
 \def\mid@dblvertical{\mskip1mu\vrule\mskip2.5mu\vrule\mskip1mu}
\fi

\usepackage{graphicx}
\usepackage[T1]{fontenc}
\usepackage[]{amsfonts}
\usepackage[]{amsmath}
\usepackage[]{rotating}
\usepackage[]{color}
\usepackage[]{float}
\usepackage[]{fancyhdr}
\usepackage[]{booktabs}
\usepackage[]{epsfig}
\usepackage{appendix}
\usepackage{amssymb}
\usepackage{psfrag}
\usepackage{setspace}
\usepackage{epstopdf}
\usepackage[normalem]{ulem}

\begin{document}

\title{Quantum Memristors with Superconducting Circuits}

\author{J. Salmilehto}
\affiliation{Department of Physics, Yale University, New Haven, Connecticut 06520, USA}
\affiliation{Department of Physical Chemistry, University of the Basque Country UPV/EHU, Apartado 644, E-48080 Bilbao, Spain}
\affiliation{QCD Labs, COMP Centre of Excellence, Department of Applied Physics, Aalto University, P.O. Box 13500, FI-00076 Aalto, Finland}
\author{F. Deppe}
\affiliation{Walther-Mei{\ss}ner-Institut, Bayerische Akademie der Wissenschaften, D-85748 Garching, Germany}
\affiliation{Physik-Department, Technische Universit\"{a}t M\"{u}nchen, D-85748 Garching, Germany}
\affiliation{Nanosystems Initiative Munich (NIM), Schellingstra{\ss}e 4, 80799 M\"{u}nchen, Germany}
\author{M. Di Ventra}
\affiliation{Department of Physics, University of California, San Diego, La Jolla, CA 92093, USA}
\author{M. Sanz} \email{mikel.sanz@ehu.eus}
\affiliation{Department of Physical Chemistry, University of the Basque Country UPV/EHU, Apartado 644, E-48080 Bilbao, Spain}
\author{E. Solano}
\affiliation{Department of Physical Chemistry, University of the Basque Country UPV/EHU, Apartado 644, E-48080 Bilbao, Spain}
\affiliation{IKERBASQUE, Basque Foundation for Science, Maria Diaz de Haro 3, 48013 Bilbao, Spain}

\begin{abstract}

Memristors are resistive elements retaining information of their past dynamics. They have garnered substantial interest due to their potential for representing a paradigm change in electronics, information processing and unconventional computing. Given the advent of quantum technologies, a design for a quantum memristor with superconducting circuits may be envisaged. Along these lines, we introduce such a quantum device whose memristive behavior arises from quasiparticle-induced tunneling when supercurrents are cancelled. For realistic parameters, we find that the relevant hysteretic behavior may be observed using current state-of-the-art measurements of the phase-driven tunneling current. Finally, we develop suitable methods to quantify memory retention in the system.

\end{abstract}

\maketitle

Circuit elements that intrinsically carry a recollection of their past evolution~\cite{pieee64/209, pieee97/1717, nano24/255201} promise to bring forth novel architectural solutions in information processing and unconventional computing~\cite{UMM} due to their passive storage capabilities. These history-dependent circuit elements can be both dissipative and non-dissipative, such as memcapacitors and meminductors~\cite{pieee97/1717, arxiv1602/07230}, or just dissipative, such as memristors. Classical memristors~\cite{ieeetct18/507, nature453/80, ejp30/661, naturenano8/13} are elements 
whose operational definition relates the voltage $V$ and the current $I$, complemented with an update of one or more internal state variable(s) $x$ carrying information of the electrical history of the system. 
For a voltage-controlled memristor
\begin{equation}
\begin{split}
I(t) &= G[x(t),V(t),t]V(t), \\
\dot{x}(t) &= f[x(t),V(t),t].
\end{split}
\end{equation}
The memductance (memory conductance) $G$ depends on both the instantaneous input voltage $V$ and the state variable $x$, which tracks the past memristor configuration via the update function $f$. Such dynamics leads to the characteristic pinched hysteresis loops under periodic driving~\cite{ieeetct18/507, pieee91/1830, ap60/145, nature453/80, nano24/255201}, a strictly non-linear conductive effect showcasing zero-energy information storage~\cite{pieee64/209}.

Even though both the quantization of superconducting circuits~\cite{devorethouches} and applications of memristors are well established techniques, memristive operation in the realm of quantum dynamics is a largely unexplored area. From an intuitive point of view, the combination of powerful memristive concepts with quantum resources, such as superposition and entanglement, promises groundbreaking advances in information and communication sciences. With this motivation in mind, the idea of a quantum memristor was recently defined in Ref.~\cite{arxiv1511/02192} by introducing the fundamental components for engineering memristive behavior in quantum systems. However, superconducting circuits naturally include memristive elements in Josephson junctions, a feature exploited in a recently proposed classical superconducting memristor design~\cite{prappl2/034011}. While this conductance asymmetric superconducting quantum interference device (CA-SQUID) design was able to produce hysteretic behavior~\cite{prappl2/034011}, it did not include the quantum features of the circuit, including the dissipative origins of the memory or its measurement and quantification. These features are of utmost importance, as the operation of the design is based on quasiparticle tunneling, whose control and measurement have recently seen significant strives forward~\cite{nature508/369, prl113/247001}. Indeed, to our knowledge, up to now no experimental work has studied the hysteretic IV-characteristics of such systems. In our opinion, the reasons for this are two-fold, namely, 1) the pinched hysteresis loops were only recently predicted to exist for such systems in Ref.~\cite{prappl2/034011} with the use of the aforementioned CA-SQUID and a proper selection of parameters, and 2) the experimental apparatus required to control and measure quasiparticle excitations with high accuracy is just beginning to emerge (see Refs.~\cite{nature508/369, prl113/247001}).

In this Article, we show that a suitably designed superconducting quantum circuit element with an external phase bias serves as a prototypical quantum memristor via low-energy quasiparticle tunneling. To this end, we describe the device in a fully quantum-mechanical fashion. We apply an ensemble interpretation of the system input and output, while the average superconducting phase difference stores information of the past dynamics. We study the hysteretic signature in a regime achievable with recent quantum nondemolition projective measurements~\cite{prl113/247001}, and construct a memory quantifier related to the accumulation of internal state change. Finally, we discuss the quantumness of our proposal, comparing it with Ref.~\cite{prappl2/034011}. Our proposal represents, to our knowledge, the first design of a superconducting quantum memristor from fundamental principles, exploiting quasiparticle tunneling in memristive quantum information processing.

\begin{figure*}[t!]
\centering
\includegraphics[width=0.95\textwidth]{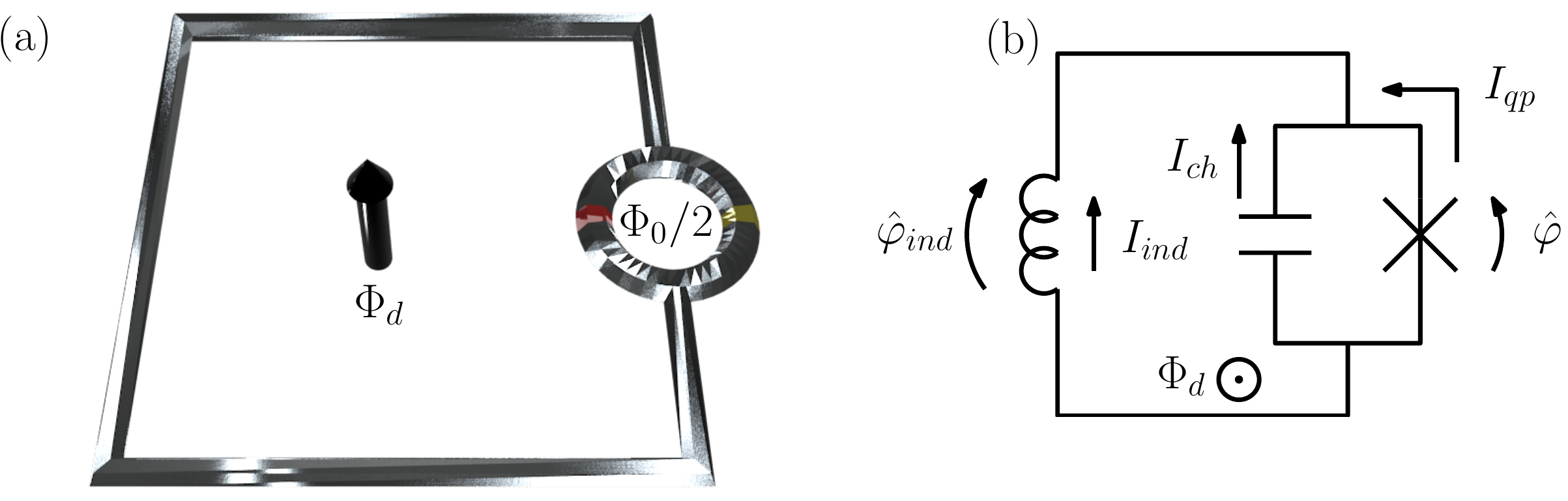}
\caption{Superconducting quantum memristor. (a) Schematic representation of the superconducting quantum memristor. The green and red strips represent junctions with different normal conductances. (b) Current diagram using effective circuit elements corresponding to the total loop inductance, charge retention of the SQUID, and the quasiparticle tunneling through it. Let us remark that, as the capacitative part has been explicitly separated in (b), the cross-notation does not refer here to the entire (effective) Josephson junction, but to the quasiparticle and phase-dependent dissipative current contributions.}
\label{fig:schematic}
\end{figure*}

The envisioned device has the rf~SQUID design shown in Fig. 1(a). It consists of a superconducting loop with inductance $L$, which is interrupted by a dc~SQUID with negligible loop inductance acting as an effective flux-tunable Josephson junction. The dc~SQUID junctions are made from different materials so that they have the same critical current but a different normal conductance~\cite{prappl2/034011}. In this way, the effective critical current of the dc~SQUID can be completely suppressed by a bias flux of half a flux quantum, $\Phi_0/2$, threading its loop~\cite{prappl2/034011}. Finally, we also apply a bias flux $\Phi_d$ to the rf~SQUID loop, resulting in the phase bias $\varphi_d = 2\pi \Phi_d/\Phi_0$. 

The total Hamiltonian of this device is the sum of the system Hamiltonian $\hat{H}_S$, a term for the quasiparticle degree of freedom, and a total tunneling term. The latter includes quasiparticle contributions but, due to the vanishing effective critical current (note that these contributions would yield a renormalization of the qubit frequency in the low-energy regime considered in Ref.~\cite{prl106/077002}), neither pair contributions, nor the Josephson counterterm~\cite{prl106/077002, prb84/064517, prb86/184514}. Under these conditions, the system Hamiltonian takes the harmonic form
\begin{equation}
\hat{H}_S = E_C \hat{n}^2 + \frac{E_L}{2} (\hat{\varphi} - \varphi_d)^2,
\label{eq:HS}
\end{equation}
where $\hat{n}$ and $\hat{\varphi}$ are the Cooper-pair counting and phase difference operators of the effective junction, respectively. We define the capacitive energy scale $E_C = 2e^2/C_d$ with  the intrinsic junction capacitance $C_d$ and the inductive energy scale $E_L = (1/L)(\Phi_0/(2\pi))^2$. Regarding the dc~SQUID, we assume the limit of strong conductance asymmetry needed for the effective junction picture due to the inclusion of quasiparticle excitations (see Supplementary Information). In this limit, the dissipative flow is through the physical junction with a smaller superconducting gap while the junction with a larger gap functions as a shunt for the total Josephson current through the SQUID. Furthermore, we demand that the phase bias is changed adiabatically, i.e., sufficiently slowly to avoid the generation of quasiparticles. Finally, our device operates in the low-energy regime $\hbar \omega_{10}, \delta E \ll 2 \Delta$, where $\omega_{10} = \sqrt{2E_C E_L}/\hbar$ is the system transition frequency and $\delta E$ is the characteristic energy of the quasiparticles above the gap $\Delta$. Even though the Hamiltonian does not warrant operation as a qubit due to the lack of sufficient anharmonicity, the system dynamics is confined to the two lowest eigenstates of Eq.~(\ref{eq:HS}) when the aforementioned assumptions are complemented with operation in the high-frequency regime $\hbar \omega_{10} \gg \delta E$. In this regime, there exist no quasiparticles with sufficiently high energy to excite the system. We emphasize that the slow biasing and high frequency assumptions utilized in this article are not contradictory. The former refers to suppressing unwanted generation of quasiparticles due to the biasing field~\cite{Zhou2014} while the latter refers to a condition on the quasiparticle bath.

The two-level master equation describing the quasiparticle-induced decay takes the Lindblad form~\cite{prb86/184514} $\partial_t \hat{\rho} = -i/\hbar [\hat{H}_S,\hat{\rho}] + \hat{\mathcal{D}}\{\hat{\rho}\}$ for the system density $\hat{\rho}$, with $\hat{\mathcal{D}}\{\hat{\rho}\}$ the corresponding Lindbladian dissipator. Note that the master equation assumes adiabatic steering, and employs the Born-Markov and secular approximations. We omit the quasiparticle-induced average frequency shift and the pure dephasing channel. See Supplemental Material for the estimation of these effects. In the low-energy limit, the decay rate factorizes into
\begin{equation}
\begin{split}
\Gamma_{1\rightarrow 0} = |\braket{0|\sin \frac{\hat{\varphi}}{2}|1}|^2 S_{qp}(\omega_{10}),
\label{eq:decay_rate}
\end{split}
\end{equation}
in the lowest order in $\omega_{10}/\Delta$. Here, $\{\ket{0},\ket{1}\}$ are the lowest energy eigenstates of $\hat{H}_S$ and the quasiparticle spectral density $S_{qp}(\omega)$ now depends on the distribution function which may, in general, includes both equilibrium and non-equilibrium contributions. Note that the decay rate in Eq.~(\ref{eq:decay_rate}) stems from the $\sin \hat{\varphi}/2 $ dependence of the quasiparticle--system coupling and is crucial to the memristive behavior detailed in the following section. By using the properties of displaced number states (see Supplementary Information), the squared inner products in Eq.~(\ref{eq:decay_rate}) have a convenient cosine form valid for any pair of Fock states $\{\ket{n},\ket{m}\}$,
\begin{widetext}
\begin{equation}
|\braket{m|\sin \frac{\hat{\varphi}}{2}|n}|^2 = \left\{
  \begin{array}{l l}
    P(g_0,n,m) [1+(-1)^{1+n-m}\cos \varphi_d]/2, & \ m\leq n \\
    P(g_0,n,m) [1+(-1)^{1+m-n}\cos \varphi_d]/2, & \ m\geq n,
  \end{array} \right.
  \label{eq:matrix_element}
\end{equation}
\end{widetext}
with
\begin{equation*}
P(g_0,n,m) = \left\{
  \begin{array}{l l}
    \exp (-g_0^2) \frac{m!}{n!} g_0^{2(n-m)} [\mathcal{L}_m^{n-m}(g_0^2)]^2, & m\leq n \\
    \exp (-g_0^2) \frac{n!}{m!} g_0^{2(m-n)} [\mathcal{L}_n^{m-n}(g_0^2)]^2, & m\geq n .
  \end{array} \right.
\end{equation*}
Here, $g_0 = [E_C/(32E_L)]^{1/4}$ and $\mathcal{L}_x^{y}$ denotes an associated Laguerre polynomial. Notably, the sign of the cosine term in Eq.~(\ref{eq:matrix_element}) depends on the parity difference between the states involved. While this potentially provides insight into interesting phenomena when multiple decay channels are involved~\cite{rpa9/35, prb10/84}, we concentrate on the two-level process and leave such considerations for future studies.

To understand how memristive behavior emerges from quasiparticle tunneling, we study the charge flow in the device. Let $\hat{a}$ be the annihilation operator for a harmonic excitation in the system. This allows us to write $\hat{\varphi} = 2 g_0 (\hat{a} + \hat{a}^{\dagger}) + \varphi_d (t)$ and $\hat{n} = i(\hat{a}^{\dagger}-\hat{a})/(4g_0)$, and denote by $\hat{\varphi}_{ind} = \hat{\varphi}-\varphi_d$ the operator for the phase over the rf~SQUID loop inductance. The directional convention for the superconducting phase differences and the different currents are presented in Fig.~\ref{fig:schematic}(b). The average charging current $\braket{\hat{I}_{ch}}$ and the inductive current $\braket{\hat{I}_{ind}}$ can be rigorously derived (see Supplementary Information) to obtain, by current conservation, the average quasiparticle current through the effective junction. The result is $\braket{\hat{I}_{qp}} =  2e\mathrm{Tr} \{\hat{\mathcal{D}}\{\hat{\rho}\}\hat{n}\} = \Gamma_{1\rightarrow 0} (-e) \braket{\hat{n}}$, which corresponds to the dissipative current induced by the interaction with the quasiparticle bath represented by the dissipator $\hat{\mathcal{D}}\{\hat{\rho}\}$. Using $\braket{\hat{V}} = -2e\braket{\hat{n}}/C_d$, the average quasiparticle current is determined by
\begin{equation}
\braket{\hat{I}_{qp}} = G_{qp}[\braket{\hat{\varphi}},\braket{\hat{V}},t] \braket{\hat{V}},
\label{eq:I_qp}
\end{equation}
where we have preemptively written the effective conductance as a function of the selected memory variable $\braket{\hat{\varphi}}$, input $\braket{\hat{V}}$, and time $t$. Solving for the dynamics, we obtain
\begin{equation*}
G_{qp}[\braket{\hat{\varphi}},\braket{\hat{V}},t] =P(g_0,1,0) S_{qp}(\omega_{10})\frac{C_d}{2}\sin^2\frac{\braket{\hat{\varphi}}-\braket{\hat{\varphi}_{ind}}}{2},
\end{equation*}
where the average inductive phase difference only requires knowledge of the input via
\begin{equation}
\begin{split}
\braket{\hat{\varphi}_{ind}} = \frac{2\pi}{\Phi_0\omega_{10}^2} \left[ \partial_t - \partial_t \ln \left(\frac{\frac{C_d g_0}{e}\braket{\hat{V}}}{\mathrm{Im} \{\rho_{01}(0) e^{i\omega_{10}t}\}}\right)\right] \braket{\hat{V}},
\label{eq:varphi_ind1}
\end{split}
\end{equation}
and we denoted the initial system coherence in the energy eigenbasis by $\rho_{01}(0) = \braket{0|\hat{\rho}|1}|_{t=0}$. The memory variable update function in $\partial_t \braket{\hat{\varphi}} = f[\braket{\hat{V}},t]$ only depends on the input and time, and has the explicit form
\begin{equation}
\begin{split}
f[\braket{\hat{V}},t] &= \frac{2\pi}{\Phi_0} \frac{1}{\omega_{10}^2} \left\{ \partial_t \ln \left(\frac{C_d g_0}{e}\frac{\braket{\hat{V}}}{\mathrm{Im} \{\rho_{01}(0) e^{i\omega_{10}t}\}}\right) \partial_t \right. \\ & \left.- \left[\partial_t \ln \left(\frac{C_d g_0}{e}\frac{\braket{\hat{V}}}{\mathrm{Im} \{\rho_{01}(0) e^{i\omega_{10}t}\}}\right)\right]^2 \right\} \braket{\hat{V}} \\ &- \frac{2e}{\hbar} \braket{\hat{V}} + \partial_t \varphi_d(t).
\label{eq:f}
\end{split}
\end{equation}
Equations (\ref{eq:I_qp})--(\ref{eq:f}) indicate that a simple superconducting device operates as a voltage-controlled quantum memristor when the average voltage over a tunneling element is interpreted as the system input, the average quasiparticle tunneling current as the output, and the average superconducting phase difference as the memory retention variable. The quasiparticle conductance acts as the memductance corresponding to the memory-dependent average current response. It should be noted that physically speaking our device is considered a flux-controlled memristive device as it includes non-zero capacitive and inductive elements \cite{pieee97/1717} while having no external capacitive coupling. However, only considering the quasiparticle contribution to the current and studying the above-mentioned equations allows us to define the device as a voltage-controlled memristor from an operational point-of-view.

The operation of the constructed memristor is of ensemble nature, that is, the system input and output are quantum averages obtained from the measurement record of the corresponding observables. Experimental input consists of initialization and a slowly oscillating flux bias applied to the rf~SQUID loop. In this way, one obtains independently generated records which, consequently, have a complex correlation exhibiting memory features via Eq.~(\ref{eq:I_qp}). In fact, the selected system input is not independent of the decay, but experiences a memory-dependent damping
\begin{eqnarray}
\braket{\hat{V}} & = & \frac{e}{C_d g_0} \exp \left[ -\frac{1}{C_d} \int_0^t G_{qp}[\braket{\hat{\varphi}},\braket{\hat{V}},\tau] d\tau \right]  \nonumber \\ 
& \times &\mathrm{Im}\{ \rho_{01}(0) e^{i\omega_{10}t} \},
\end{eqnarray}
which allows one to self-consistently solve the fundamental equations above. One such solution is identifiable as mimicking the operation of the classical superconducting memristor~\cite{prappl2/034011}, in which the memory is fully stored in the phase bias. It is obtained in the weak-damping limit by initializing the system with $\braket{\hat{V}}|_{t=0} = V_0$ and $\braket{\hat{\varphi}}|_{t=0} = \varphi_d(0)$, and by assuming a resonant sinusoidal phase bias $\varphi_d (t) = \varphi_0 + (2eV_0)/(\hbar \omega_{10}) \sin (\omega_{10} t)$. Weak voltage damping implies that $\braket{\hat{V}} \approx  e/(C_d g_0) \mathrm{Im}\{ \rho_{01}(0) e^{i\omega_{10}t} \} = V_0 \cos (\omega_{10}t)$, where the update is given by the classical Josephson relation $\partial_t \varphi_d(t) = 2e/\hbar \braket{\hat{V}}$. The solution embodies the two implicit assumptions for the classical memristor: (1) the rf~SQUID loop has a negligible inductance, and (2) the internal dynamics is negligibly affected by the same dissipation that produces the output.

\begin{figure}
\centering
\includegraphics[width=0.5\textwidth]{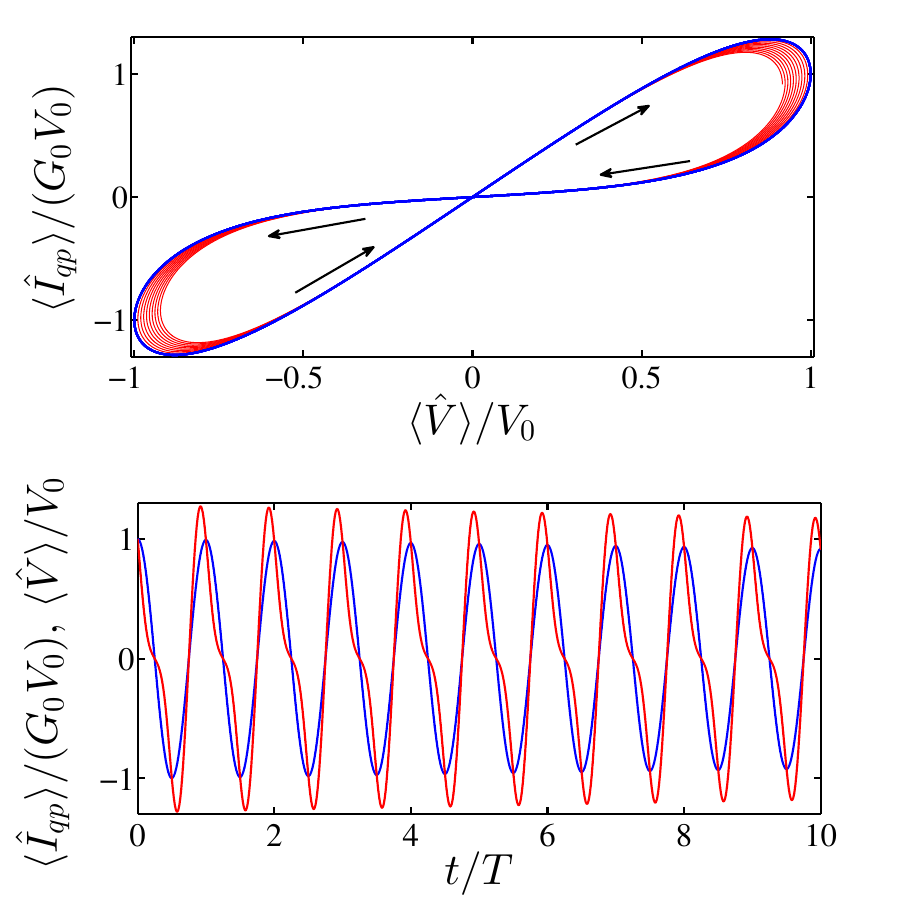}
\caption{Left panel: Parametric hysteresis curve (red) and the weak-damping solution (blue). Rigth panel: Temporal evolution of average voltage (blue) and quasiparticle current (red). We use initialization $\braket{\hat{V}}|_{t=0} = V_0$, $\braket{\hat{\varphi}}|_{t=0} = \varphi_d(0)$ and resonant sinusoidal phase bias. Parameters are $E_C/(2\pi\hbar)=1$ GHz, $E_L=10^3E_C$, $\varphi_0 = \pi/2$, $S_{qp}(\omega_{10}) = \omega_{10}$, and $2eV_0/(\hbar \omega_{10}) = 1$. The arrows indicate the direction of temporal evolution recorded over 10 oscillation periods $T=2\pi/\omega_{10}$ and we use the shorthand notation $G_0 = P(g_0,1,0)S_{qp}(\omega_{10}) C_d/4$.}
\label{fig:hysteresis}
\end{figure}

As a first step, we need to verify whether the above-described classical-limit solution is consistent with the semiclassical results of Ref.~\onlinecite{prappl2/034011}. In Fig~\ref{fig:hysteresis}, one clearly sees that we observe the hysteretic current-voltage characteristic curves as required for a memristive element. In other words, a proper choice of the sinusoidal drive allows for tunable finite-area pinched loops~\cite{nano24/255201}. Employing the system parameters from Fig.~\ref{fig:hysteresis}, the above weak-damping solution is accurately numerically retrieved with $S_{qp}(\omega_{10}) = 10^{-4} \omega_{10}$ over multiple oscillation periods. This corresponds to a minimum relaxation time during the driving period of $\min ( T_1 ) \propto 1 $~$\mu$s relevant to the current state-of-the-art experimental setups~\cite{nature508/369, prl113/247001}. While those setups consider a different type of system, the fluxonium, very little experimental work has been able to reach the regime in which quasiparticle-induced relaxation is observable and, consequently, we use these references for initial comparison. Even though $\braket{\hat{I}_{qp}} \propto S_{qp}(\omega_{10})$, the magnitude-scaled hysteresis curve is robust against decreasing the minimum $T_1$-time by 2 orders of magnitude (see Supplementary Information). Beyond this, the input and output values are subject to noticeable decay. We show the parametric dependence of the average voltage and quasiparticle current in Fig.~\ref{fig:hysteresis} for $S_{qp}(\omega_{10}) = \omega_{10}$ corresponding to $\min ( T_1 ) \propto 100$~ps.
The hysteresis curve starts from a point in the weak-damping trajectory due to the identical initialization, and it is followed by a reduction of the area with time. The time evolution in Fig.~\ref{fig:hysteresis} shows a gradual decay in the voltage and current amplitudes. Note that the system is operated in the phase regime of almost negligible loop inductance. This allows for a feasible resonant phase biasing frequency $\omega_{10}/2\pi \approx 45$~GHz, achieved while ensuring sufficient adiabaticity $\max (\alpha_{rs}) \approx 0.15$~(see Supplementary Information), necessary for the master-equation treatment employed for the quasiparticle bath.

\begin{figure}
\centering
\includegraphics[width=0.5\textwidth]{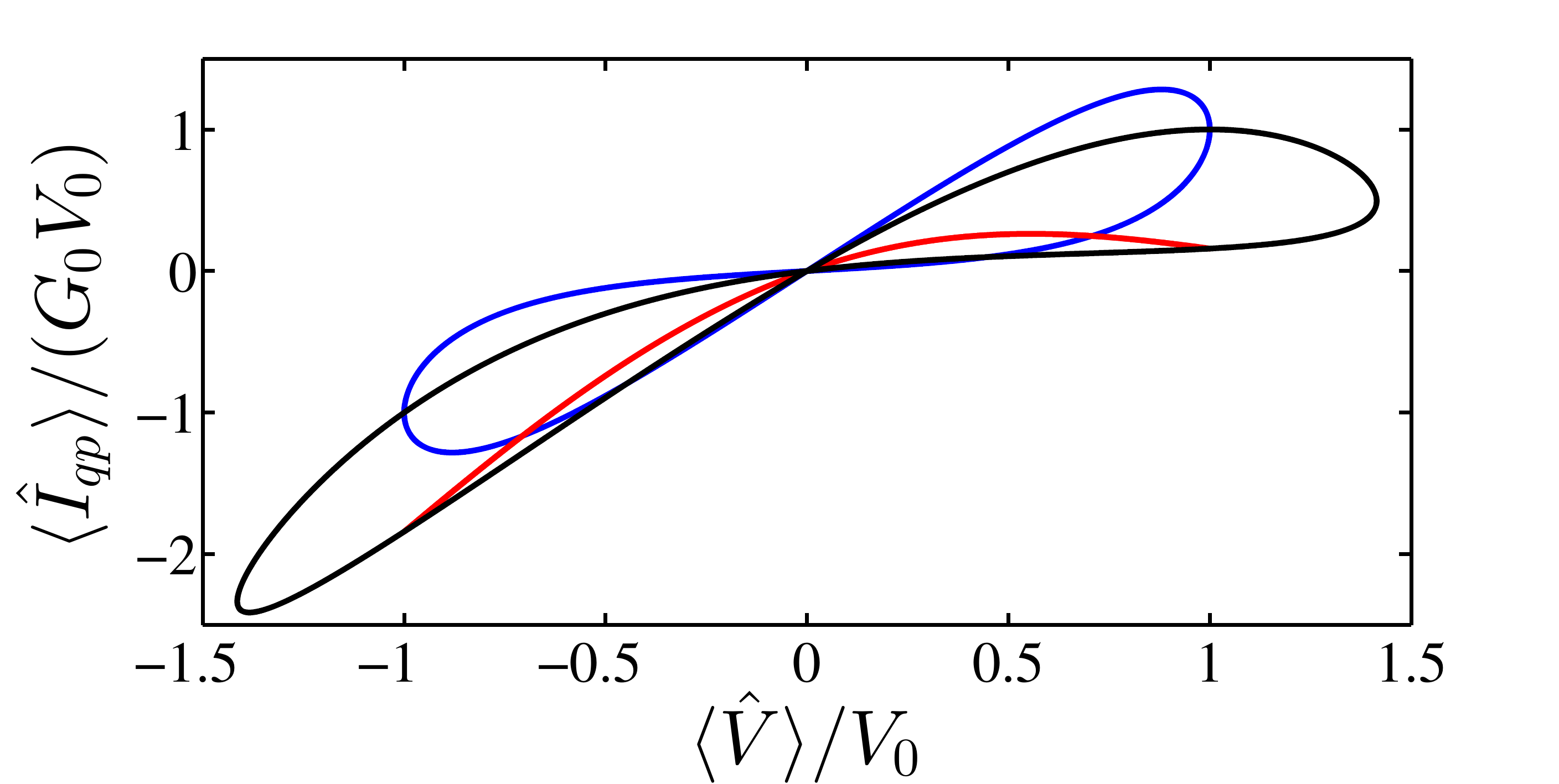}
\caption{Hysteresis curves with resonant sinusoidal phase bias in the weak-damping limit. System initialized such that $\braket{\hat{V}}|_{t=0} = V_0$, $\braket{\hat{\varphi}}|_{t=0} = \varphi_d(0)$ (blue), $\braket{\hat{V}}|_{t=0} = 0$, $\braket{\hat{\varphi}}|_{t=0} = 4g_0^2 V_0 C_d/e + \varphi_d(0)$ (red), and $\braket{\hat{V}}|_{t=0} = V_0$, $\braket{\hat{\varphi}}|_{t=0} = 4g_0^2 V_0 C_d/e + \varphi_d(0)$ (black). The system parameters are the same as in Fig.~\ref{fig:hysteresis} with $S_{qp}(\omega_{10}) = 10^{-4} \omega_{10}$.}
\label{fig:init}
\end{figure}

The initialization of the system plays a crucial role in the operation and does not simply determine the initial position in the parametric curve. Figure~\ref{fig:init} shows the hysteresis curves for three different initializations, assuming the weak-damping limit and a resonant sinusoidal drive protocol. These curves can be interpreted by studying the time symmetry of the quasiparticle current between two consecutive crossings of the zero-energy point and indicate a tunable landscape of hysteretic behavior (see Supplementary Information).

To quantify the non-Markovian~\cite{njp13/093004} character of the device, we consider the area enclosed by a hysteresis loop in the current-voltage plane as a memory measurement. This interpretation is founded in the observation that the absence of area correlates with purely time-local current response. In other words, a nonlinear conductance cannot produce a non-zero area since it depends only on the instantaneous value of the voltage. The memory quantifier for the $k$th traversed loop takes the form $N_m^k = \int_{t_c^k}^{t_c^{k+1}} dt F_m(t)$ (see Appendix \ref{area}), where $t_c^k$ fulfills $\braket{\hat{V}}|_{t=t_c^k} = 0$ for each $k$. This quantifier stores the evolution information of $F_m(t) = F_0(t) + \braket{\hat{V}}^2 \partial_t G_{qp}[\braket{\hat{\varphi}},\braket{\hat{V}},t]/2$, where 
\begin{eqnarray}
F_0(t) &= f[\braket{\hat{\varphi}},\braket{\hat{V}},t] \frac{1}{2} \braket{\hat{V}}^2 \partial_{\braket{\hat{\varphi}}} G_{qp}[\braket{\hat{\varphi}},\braket{\hat{V}},t] \nonumber \\ &= \Delta \braket{\hat{\varphi}} \frac{1}{2} \frac{d}{dt} \left(-\braket{\hat{V}}^2 \partial_{\braket{\hat{\varphi}}}G_{qp}[\braket{\hat{\varphi}},\braket{\hat{V}},t]\right)
\label{eq:F_0}
\end{eqnarray}
corresponds to the response specific to the selected memory variable, and the second term to the explicit time dependence of the memductance not included in the internal memory variable. However, it is in principle always possible to redefine the memory variable to absorb the explicit time dependence in the memductance, so that $F_m(t) = F_0(t)$. The two expressions in Eq.~(\ref{eq:F_0}) imply that the quantifier corresponds to a time-dependent weighted record of the change in the memory variable $\partial_t \braket{\hat{\varphi}}$ or the instantaneous distance from its initial value $\Delta \braket{\hat{\varphi}} = \braket{\hat{\varphi}} - \braket{\hat{\varphi}}|_{t=t_c^k}$. If the conductance is a non-linear function of only the instantaneous input, $F_m(t)$ vanishes in integration due to input periodicity. See Supplemental Information for the decay of the quantifier as well as its response to different initializations.

Finally, our quantum memristor is formulated in the ideal case of zero leakage supercurrent. Adding a nonzero pair-tunnelling term, not only modifies the energy and state structure, but inflicts a Josephson tunneling current which may disrupt the operation. While there can be multiple factors contributing to the leakage supercurrent, such as magnetic flux noise, the primary experimental factor to tackle is possibly the critical current imbalance in the SQUID. The state-of-the-art critical current suppression factor based purely on fabrication techniques is $\sim$$10^{-2}$ while the balanced SQUID~\cite{apl92/052110} promises a factor of $\sim$$10^{-3}$--$10^{-4}$, for a maximum critical current of 30 nA. In terms of the Hamiltonian, this implies that the imbalance term is $10^{-1}$--$10^{-3}$ times the charging energy scale used here. In addition, our formulation assumes only the quasiparticle decay channel and omits other natural loss channels (dielectric, inductive, radiative). Recent experimental work has studied quasiparticle-limited relaxation and shown significant progress in suppressing the additional decay channels, modifying the quasiparticle population through different means, and discerning between the different decay mechanisms~\cite{nature508/369, prl113/247001}.

Let us finish with a brief discussion about the quantumness of the system, as well as the role of superposition and entanglement. The dynamics of the quantum memristor described above is purely quantum, in the strict sense that the evolution cannot be emulated by a classical channel~\cite{wielandt,2013arXiv1312.1329I}. This is not surprising, since the quantumness of our design refers to the full dissipative treatment (as an open quantum system) of the quasiparticle bath leading to memristive features in the expectation values of quantum observables. Therefore, superposition plays the same role as in any other quantum system. With respect to the entanglement, coupling two of these quantum memristors is a natural and relevant question after showing the dynamics of a single device, but beyond the scope of this manuscript.

In conclusion, we have demonstrated a prototype design for a quantum memristor in a superconducting circuit relying on quasiparticle tunneling. The pinched hysteretic behavior of the average quasiparticle current is a clear signature of conductance beyond typical non-linearity, and modified by both the characteristics of the circuit and the quasiparticle bath. The measurement resolution can potentially be varied by tuning the non-equilibrium quasiparticle population, by just using the state-of-the-art injection and trapping methods~\cite{prl113/247001} during the lifetime of the quasiparticles. Our work paves the way for the engineering of on-demand quantum non-Markovianity using the superconducting quantum memristor as a building block. Furthermore, we may consider possible applications such as the codification of Quantum Machine Learning protocols~\cite{arxiv1507/08482, arxiv1307/0411} and neuromorphic quantum computing~\cite{arxiv1511/02192}.

 \section*{AUTHOR CONTRIBUTIONS}
J.S., as the first author, has been responsible for the development of this work. M.S. supported J.S. with the mathematical demonstrations and calculations. M.S. and E.S. suggested the seminal ideas. F.D and M.D. has helped to check the feasibility and improve the ideas and results shown in the paper. All authors have carefully proofread the manuscript. E.S. supervised the project throughout all stages.

\section*{ACKNOWLEDGMENT}
The authors would like to thank Steven Girvin, Paul Pfeiffer, and Mikko M\"ott\"onen for helpful discussions. J.S., F.D., M.S., and E.S. aknowledge finantial support from the CCQED EU project, J.S. ackowledges the Finnish Cultural Foundation, M.S. and E.S. aknowledge the support from MINECO/FEDER Grant No. FIS2015-69983-P, Basque Government Grant No. IT986-16, and UPV/EHU UFI 11/ 55, and M. D. acknowledges support from the DOE under grant DE-FG02-05ER46204. 
\section*{ADDITIONAL INFORMATION} 
The authors declare no competing financial interests.

\appendix

\section{The effective junction picture in the presence of quasiparticle excitations}

While it is well-known that a SQUID behaves as if it was a single effective junction with a tunable Josephson energy when pair tunneling is discussed, additional conductance requirements are set by the inclusion of quasiparticle excitations. Classically speaking, a non-vanishing phase-dependent conductance can be achieved in a conductance-asymmetric SQUID, while assuming that the physical junctions have equivalent critical currents~\cite{prappl2/034011}. Summation of dissipative currents yields a representation as an effective junction with an effective leakage conductance $G_{\mathrm{eff}} = G_1 + G_2$, where $G_i$ is the leakage conductance of the $i$th physical junction, and an asymmetry term for the phase-dependent current of $S_{\mathrm{asym}} = (G_1 - G_2)/(G_1+G_2)$~\cite{prappl2/034011}. 

To use the effective junction picture in the main text in association with quasiparticle tunneling, we assume strong asymmetry $G_1 \gg G_2$ such that $S_{\mathrm{asym}} \approx 1$. Hence, the dissipative flow is effectively only through a single junction, while the total pair current can still be cancelled. This allows for the effective junction picture to be used for describing quasiparticle tunneling. Since the Ambegaokar-Baratoff relation implies for equivalent critical currents that $G_1/G_2 = \Delta_2/\Delta_1$, with $\Delta_i$ is the superconducting gap of the electrodes of the $i$th physical junction of the SQUID, this assumption can be enforced by demanding $\Delta_2 \gg \Delta_1$. Weakening the assumption would require the degrees of freedom of the individual physical qubits to be included in the Hamiltonian in Eq.~(2) of the manuscript. This means that the quasiparticle tunneling Hamiltonian would have to be given for each physical junction separately, resulting in two separate decay channels, which allows for the weaker conductance discrepancy to manifest. We leave such considerations for future work and assume the limit of strong conductance asymmetry throughout the main text.

\section{Estimation of adiabaticity, quasiparticle-induced average frequency shift, and pure dephasing}

\textit{Adiabaticity.}-- In order to evaluate the adiabaticity of our phase-driven system described by the Hamiltonian in Eq.~(1) of the main text, we calculate the instantaneous adiabatic parameter for the dynamics confined to the two lowest energy levels. The instantaneous eigenstates of the abovementioned Hamiltonian are the well-known instantaneous number states 
\begin{eqnarray}
\ket{n(t)} & = & (2^n n! \sqrt{\pi} d_0)^{-\frac{1}{2}} \int d\varphi \exp\left[ -\frac{1}{2} \left( \frac{\varphi - \varphi_d(t)}{d_0} \right)^2 \right]  \nonumber \\
& \times & H_n \left( \frac{\varphi - \varphi_d(t)}{d_0} \right),
\label{eq:nstate}
\end{eqnarray}
where $d_0 = (2E_L/E_C)^{\frac{1}{4}}$ and $H_n$ is the $n$th Hermite polynomial. By using known properties for the Hermite polynomials, we obtain two useful identities
\begin{equation} \label{hermite1}
\partial_t H_n \left( \frac{\varphi - \varphi_d(t)}{d_0} \right) = -\frac{\partial_t \varphi_d(t)}{d_0} \partial_n H_{n-1} \left( \frac{\varphi - \varphi_d(t)}{d_0} \right),
\end{equation}
and
\begin{eqnarray} \label{hermite2}
\frac{\varphi-\varphi_d(t)}{d_0} H_n \left( \frac{\varphi - \varphi_d(t)}{d_0} \right) & = & \frac{1}{2} H_{n+1} \left( \frac{\varphi - \varphi_d(t)}{d_0} \right) \nonumber \\
& + & n H_{n-1}\left( \frac{\varphi - \varphi_d(t)}{d_0} \right).
\end{eqnarray}
By taking a time-derivative of the instantaneous eigenstate in Eq.~(\ref{eq:nstate}) and applying the identities given by Eqs. \eqref{hermite1} and \eqref{hermite2}, we obtain
\begin{equation}
\braket{m(t)|\partial_t|n(t)} = \frac{\partial_t \varphi_d(t)}{\sqrt{2}d_0} \left(\sqrt{n+1} \delta_{m,n+1}-\sqrt{n}\delta_{m,n-1}\right),
\label{eq:derinner}
\end{equation}
where the remaining time-derivative depends on the details of the external drive protocol.

For the dynamics confined to the two lowest levels, the instantaneous adiabatic parameter is $\alpha_{\text{adi}} = ||\hat{w}||/\omega_{10}$, where $||\hat{w}|| = \mathrm{Tr} \{ \hat{w}^{\dagger} \hat{w} \} ^{\frac{1}{2}}$ and $\hat{w} = -i \hat{D}_w^{\dagger}\partial_t\hat{D}_w$ generates the Berry connection. Here, $\hat{D}_w = \ket{0(t)}\bra{0_f} + \ket{1(t)}\bra{1_f}$ and $\{ \ket{0_f},\ket{1_f}\}$ is an orthonormal diabatic basis. By using Eq.~(\ref{eq:derinner}), the adiabatic parameter takes a simple form
\begin{equation}
\alpha_{\text{adi}} = \frac{|\partial_t \varphi_d(t)|}{\omega_{10}d_0} = \frac{\hbar |\partial_t \varphi_d(t)|}{(8E_CE_L^3)^{\frac{1}{4}}}.
\end{equation}

For the resonant sinusoidal driving $\varphi_d(t) = \varphi_0 + \frac{2eV_0}{\hbar \omega_{10}} \sin (\omega_{10} t)$ used in the simulations in the main text, the adiabatic parameter becomes
\begin{equation}
\alpha_{rs} = \frac{2eV_0}{\hbar \omega_{10} d_0} |\cos (\omega_{10}t)| = \frac{2e V_0}{(8E_CE_L^3)^{\frac{1}{4}}} |\cos (\omega_{10}t)|.
\end{equation}
In order for the master equation approach in the main text to be valid, Landau-Zener transitions must be suppressed at all times, such that any population transfer is dissipation--induced and occurs between instantaneous eigenstates. This is guaranteed by operating in the regime $\max (\alpha_{rs}) \ll 1$.

\textit{Quasiparticle-induced average shift of the system transition frequency.}-- The existence of quasiparticles induces an average frequency shift for the system that can be attributed to two different mechanisms~\cite{prb84/064517, prb86/184514}: the quasiparticle renormalization of the Josephson energy and the quasiparticle-mediated virtual transitions between different energy levels. In general, each physical junction generates different shift terms, since nonvanishing total quasiparticle current requires conductance asymmetry (see previous section) and, hence, different superconducting gaps for the electrodes of the junctions ($\Delta_1 \neq \Delta_2$). However, our assumption of strong conductance asymmetry ($\Delta_1 \ll \Delta_2$) implies that the frequency shift is dominated by the junction with large quasiparticle flow, so we denote the effective gap by $\Delta \approx \Delta_1$. The renormalization term comprises of contributions from the pair tunneling and Josephson counterterms ($\hbar \omega_{10},\delta E \ll 2\Delta$), as well as the terms in the quasiparticle tunneling Hamiltonian that do not contribute to pure dephasing and relaxation~\cite{prb86/184514}. Since these contributions are renormalized Josephson energy terms for the individual junctions, they vanish for our effective junction. For a nonvanishing Josephson term, the quasiparticle contribution can be approximated knowing the CP-density-normalized quasiparticle density $x_{qp}$ and the energy mode occupation of the quasiparticles at the gap $x_{qp}^A = f_{E,qp}(\Delta)$, where $f_{E,qp}$ is the energy mode of the lead quasiparticle distribution function~\cite{prb84/064517}.

By using the assumptions detailed in the main manuscript, the principal contribution from the virtual transitions to the $i$th energy level of the system is~\cite{prb84/064517}
\begin{equation}
\delta E_{i,qp} = \sum_{k\neq i} \left| \braket{k|\sin \frac{\hat{\varphi}}{2}|i}\right|^2 F_{qp}(\omega_{ki}),
\label{eq:qp_shift}
\end{equation}
where $\ket{i}$ is the $i$th eigenstate with energy $E_i$,  $\hbar \omega_{ki} = E_k - E_i$, and $F_{qp}(\omega)$ is an expression involving complex nested integration of the quasiparticle distribution in energy space [see Appendix A of Ref.~\cite{prb84/064517}]. As it is apparent from Eq.~(\ref{eq:qp_shift}), each energy shift generally accounts for virtual transitions to and from each other state in the Hilbert space. Our system lacks anharmonicity and, hence, we cannot restrict virtual occupation to the two lowest levels. However, we operate in the phase regime where $g_0 = [E_C/(32E_L)]^{1/4} \ll 1$ and, hence, Eqs.~(4) and (5) in the main text imply that the largest terms in $\delta E_{i,\mathrm{qp}}$ correspond to virtual transitions between $i$ and its energetically nearest levels. Thus, the frequency shift becomes
\begin{eqnarray}
\hbar \delta \omega_{10,qp} & = & \delta E_{1,qp} - \delta E_{0,qp} \nonumber \\
& = &g_0^2 \frac{1 + \cos\varphi_d}{2} [F_{qp}(\omega_{10}) + F_{qp}(-\omega_{10})]
\label{eq:qp_shift2}
\end{eqnarray}
in order $O(g_0^4)$, where the non-nearest-neighbour terms are of the order $g_0^4$ by the constuction of the inner product in Eq.~(\ref{eq:qp_shift}), whilst the nearest-neighbour terms scale as $e^{-g_0^2}g_0^2$, which was expanded to obtain the principal term in Eq.~(\ref{eq:qp_shift2}). Notably, the lowest-order contributions stem from the virtual transitions $1\leftrightarrow 0$, $0\leftrightarrow 1$, and $2\leftrightarrow 1$. Making use of the definition of quasiparticle impedance and assuming the high-frequency limit, we finally obtain
\begin{equation}
\hbar \delta \omega_{10,qp} = -  g_0^2 \frac{\hbar g_T \Delta}{e^2} \left[ x_{qp} \sqrt{\frac{2\Delta}{|\hbar \omega_{10}|}} - 2\pi x_{qp}^A \right] \sin^2\frac{\varphi_d}{2}
\label{eq:qp_shift3}
\end{equation}
in order $O(g_0^4)$where $g_T \propto G_{\mathrm{eff}}$ is the effective junction conductance and $e$ is the elementary charge. This term is generally time-dependent, due to the phase steering. We assume that the physical parameters are selected such that $\delta \omega_{10,qp} \ll \omega_{10}$, so that the frequency shift can be omitted. As evident from Eq.~(\ref{eq:qp_shift3}), the validity of this assumption is determined by the details of the execution of the low--energy limit given in the main text hindering it and the execution of the phase--regime as well as typical low quasiparticle densities supporting it. Note that in thermal equilibrium and at low temperatures $T \ll \Delta$, it is straightforward to support our statement that the frequency shift is dominated by the junction with large dissipative flow ($\Delta_1 \ll \Delta_2$) using Eq.~(\ref{eq:qp_shift3}), since then $\delta \omega_{10,qp} \propto e^{-\Delta /k_B T}$. This implies that $\delta \omega_{10,qp}^1/\delta \omega_{10,qp}^2 \propto e^{(\Delta_2 -\Delta_1) /k_B T} \gg 1$, where the superscript indicates the physical junction.

\textit{Pure dephasing.}-- For our system, whose memristive operation relies solely on a weak but nonvanishing dissipative current generated by the quasiparticle--induced decay, pure dephasing adds another decoherence channel and, might potentially destroy the memristive behavior. Its effect would be to both distort the hysteresis loops by adding another time-dependent contribution to the quasiparticle current and to increase the memory-dependent damping of the average voltage by contributing to the total dephasing. Even though our SQUID is necessarily asymmetric as explained earlier, the individual junctions are assumed symmetric throughout this work. Thus, one would estimate pure dephasing using the self-consistent rate $\Gamma_{\phi}$ in Eq.~(28) of  Ref.~\cite{prb86/184514} to avoid the issue of logarithmic divergence in the lowest-order perturbative tunneling treatment. However, the self-consistent rate scales with the system energy as $|A_s^d|^2$ where $A_s^d = (\braket{1|\sin \hat{\varphi}/2|1} - \braket{0|\sin \hat{\varphi}/2|0})/2$ and the inner products can be calculated using the identities in Eqs.~(\ref{eq:sinvarphi}) and (\ref{eq:displacement}). This yields $|A_s^d|^2 = g_0^4e^{-g_0^2}\sin^2 (\varphi_d/2)/4$, which is of the order $g_0^4$, while the relaxation rate scales as $g_0^2$. A full comparison of the relaxation and pure dephasing rates would require knowledge of the quasiparticle distribution to calculate the quasiparticle spectral density in Eq.~(3) of the main text, as well as to iteratively solve the self-consistent expression for the pure dephasing rate. As an example, assuming $\Gamma_{\phi} \ll \delta E$, the scaling becomes $\Gamma_{\phi} \propto |A_s^d|^2 \ln (1/|A_s^d|^2)$~\cite{prb86/184514}, which decreases faster than $\Gamma_{1\rightarrow 0}$ in small $g_0$. In this work, we assume that the quasiparticle distribution can be established in a manner that the beneficial difference in scaling in the phase regime allows for the pure dephasing process to be neglected.

Finally, we remind the reader that our construction of the relaxation rate, as well as the considerations on the pure dephasing rate above, do not only exploit the adiabatic assumption but also that of low charateristic quasiparticle energies ($\hbar \omega_ {10}, \delta E  \ll 2 \Delta$). This implies that the \textit{secondary} terms in the rate equations in Ref.~\cite{prb86/184514} can typically be omitted, since they are negligible in comparison to the primary terms we use throughout this work. However, our system crosses points during the phase-driving process in which the primary terms vanish and the secondary terms become the dominating contributions, ensuring that the total rates are nonvanishing. For example, the secondary term in the relaxation rate is proportional to $e^{-g_0^2}g_0^2 (1-\cos \varphi_d)/2$, which reaches its maximum value at the point of vanishing primary term proportional to $e^{-g_0^2}g_0^2 (1+\cos \varphi_d)/2$, that is, at $\varphi = \pi + 2\pi n$, $n \in \mathbb{Z}$. Even though the secondary terms dominate near these points, we assume that their contribution to the total dissipative flow during the full driving cycle is negligible due to the aforementioned assumptions. If low characteristic quasiparticle energies cannot be guaranteed, the full rate equations should be applied to study the potentially modified memristive function.

\section{Matrix element of $\sin \hat{\varphi}/2$ in the quasiparticle decay rate}

By using $\hat{\varphi} = 2 g_0 (\hat{a} + \hat{a}^{\dagger}) + \varphi_d (t)$, where $\hat{a}$ is the bosonic annihilation operator for the harmonic system, the sinusoidal phase term can be written as
\begin{equation}
\sin \frac{\hat{\varphi}}{2} = \frac{1}{2i} \left[e^{i\varphi_d/2}\hat{D}(i g_0)  - e^{-i\varphi_d/2}\hat{D}(-ig_0)\right],
\label{eq:sinvarphi}
\end{equation}
where $\hat{D}(ig_0) = \exp[ig_0(\hat{a}+\hat{a}^{\dagger})]$ is the displacement operator with $ig_0 = i[E_C/(32E_L)]^{1/4}$ the phase space displacement. General properties for displaced number states assert that, for any pair of eigenstates of the harmonic oscillator $\ket{n}$ and $\ket{m}$, we have the identity~\cite{prd19/1669}
\begin{equation}
\braket{m|\hat{D}(\alpha)|n} = \left\{
  \begin{array}{l l}
    e^{\frac{-|\alpha|^2}{2}} \sqrt{\frac{m!}{n!}} (-\alpha^*)^{n-m} \mathcal{L}_{m}^{n-m} (|\alpha|^2), & m\leq n \\
   e^{\frac{-|\alpha|^2}{2}} \sqrt{\frac{n!}{m!}} \alpha^{m-n} \mathcal{L}_{n}^{m-n} (|\alpha|^2), & m\geq n,
  \end{array} \right.
  \label{eq:displacement}
\end{equation}
where $\alpha$ is an arbitrary displacement and $\mathcal{L}_x^{y}$ denotes an associated Laguerre polynomial. The identity for $\alpha = ig_0$ results in
\begin{widetext}
\begin{equation}
\braket{m|\hat{D}(ig_0)|n} \braket{m|\hat{D}(-ig_0)|n}^* = \braket{m|\hat{D}(-ig_0)|n} \braket{m|\hat{D}(ig_0)|n}^* = \left\{
  \begin{array}{l l}
    (-1)^{n-m} |\braket{m|\hat{D}(ig_0)|n}|^2, & \ m\leq n \\
    (-1)^{m-n} |\braket{m|\hat{D}(ig_0)|n}|^2, & \ m\geq n.
  \end{array} \right.
  \label{eq:Dig0ident}
\end{equation}
\end{widetext}
Application of Eqs.~(\ref{eq:sinvarphi}) and (\ref{eq:Dig0ident}) yields Eqs.~(3) and (4) in the main text. It should be noted that a similar calculation has been performed in Ref.~\cite{prb84/064517} for the pair ($\ket{n},\ket{0}$) and the results presented in the main text are a generalization to any pair of Fock states. The squared inner products corresponding to the transitions between the three lowest energy eigenstates are
\begin{equation}
\begin{split}
|\braket{0|\sin \frac{\hat{\varphi}}{2}|1}|^2 &= e^{-g_0^2}g_0^2\frac{1+\cos\varphi_d}{2}, \\
|\braket{0|\sin \frac{\hat{\varphi}}{2}|2}|^2 &= \frac{1}{2}e^{-g_0^2}g_0^4\frac{1-\cos\varphi_d}{2}, \\
|\braket{1|\sin \frac{\hat{\varphi}}{2}|2}|^2 &= \frac{1}{2}e^{-g_0^2}g_0^2(2-g_0^2)^2\frac{1+\cos\varphi_d}{2}.
\end{split}
\end{equation}

\section{Derivation of the charging, inductive, and quasiparticle currents}

The average charging current for the junction is given by
\begin{equation}
\begin{split}
\braket{\hat{I}_{ch}} = -2e \partial_t \braket{\hat{n}} = \frac{2e}{\hbar}E_L \mathrm{Tr} \{ \hat{\rho}(\hat{\varphi}-\varphi_d) \}  -  2e\mathrm{Tr} \{\hat{\mathcal{D}}\{\hat{\rho}\}\hat{n}\},
\end{split}
\end{equation}
where we assumed a Lindblad-form evolution $\partial_t \hat{\rho} = -i/\hbar [\hat{H}_S,\hat{\rho}] + \hat{\mathcal{D}}\{\hat{\rho}\}$ for the system density $\hat{\rho}$, and used the properties of the bosonic operators to rewrite the commutator $[\hat{n},\hat{H}_S] = -i\hbar \omega_{10} (\hat{a}^{\dagger}+\hat{a})/(4g_0)$, after applying Ehrenfest theorem. By employing the operator for the current through the inductive element $\hat{I}_{ind} = -2e/\hbar E_L (\hat{\varphi} - \varphi_d)$, the average inductive current is
\begin{equation}
\begin{split}
\braket{\hat{I}_{ind}} =\mathrm{Tr} \{ \hat{\rho} \hat{I}_{ind} \} = -\frac{2e}{\hbar}E_L \mathrm{Tr} \{ \hat{\rho}(\hat{\varphi}-\varphi_d) \}.
\end{split}
\end{equation}
Hence, current conservation dictates that $\braket{\hat{I}_{qp}} = - \braket{\hat{I}_{ch}} - \braket{\hat{I}_{ind}} =  2e\mathrm{Tr} \{\hat{D}\{\hat{\rho}\}\hat{n}\}$. We assume the high-energy low-frequency regime discussed in the main text, so that $\hat{\mathcal{D}}\{\hat{\rho}\} = \hat{L} \hat{\rho} \hat{L}^{\dagger} - \{\hat{L}^{\dagger}\hat{L},\hat{\rho} \}/2$, where $\hat{L} = \sqrt{\Gamma_{1 \rightarrow 0}} \hat{a}$. Decomposing $\hat{\mathcal{D}}\{\hat{\rho}\} \hat{n}$ inside the trace and using the properties of $\hat{a}$ yields $\braket{\hat{I}_{qp}} = \Gamma_{1 \rightarrow 0} (-e) \braket{\hat{n}}$, given in the main text.

\section{Decay of hysteresis with respect to the scaling of $T_1$-times}

\begin{figure}
\centering
\includegraphics[width=0.5\textwidth]{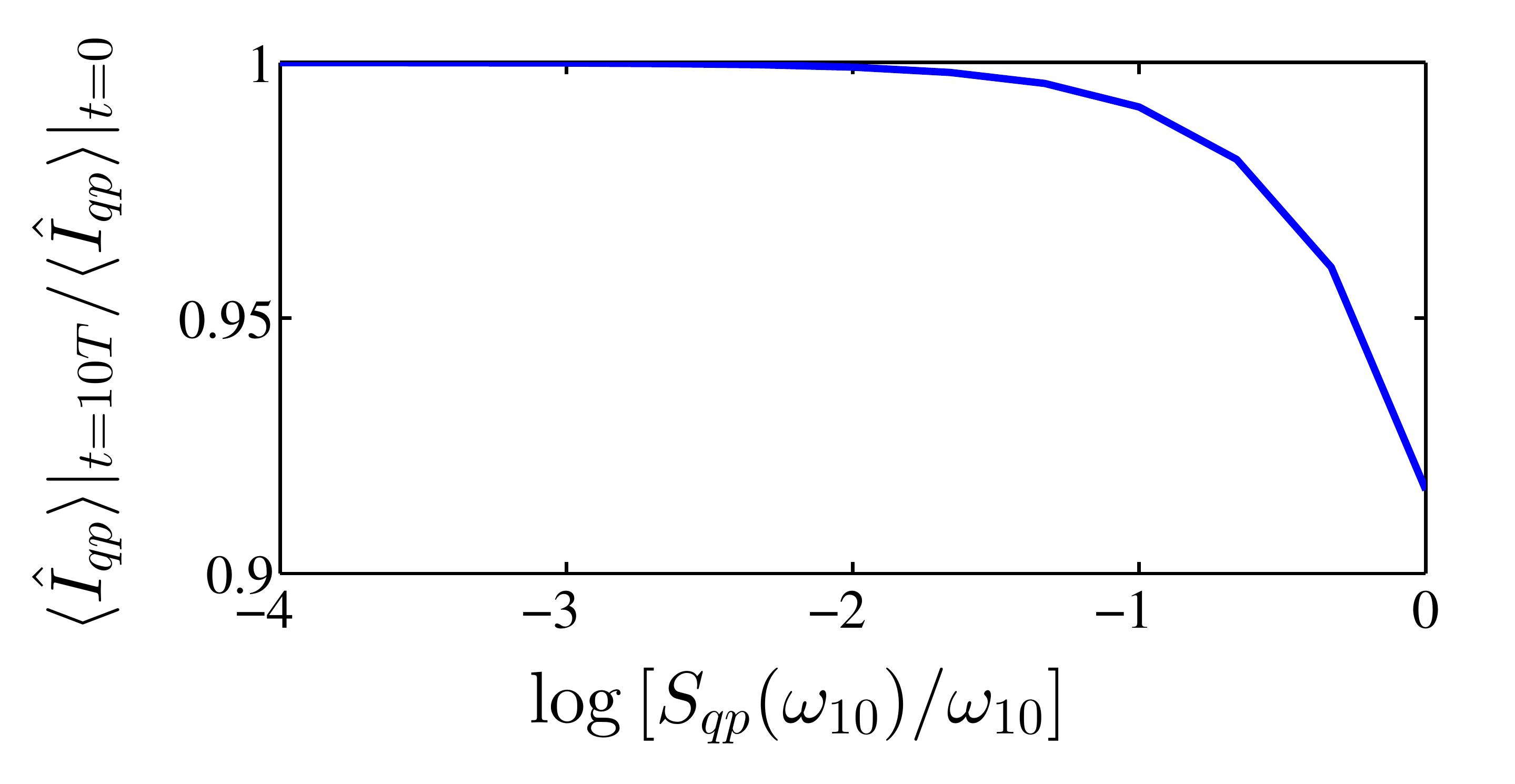}
\caption{Expectation value of the quasiparticle current after 10 resonant sinusoidal driving cycles with respect to the spectral density of the quasiparticle bath. System parameters and initialization as in Fig. 2 of the main text.}
\label{fig:scaling}
\end{figure}

The same decay that enables hysteresis is responsible for the gradual decrease of the quasiparticle current due to dephasing. Observation of hysteresis requires a sufficently large $T_1$-time to be achieved. To estimate the bath-induced decay of the input and output values with respect to the scaling of the $T_1$-times, we present the change in the average quasiparticle current after 10 consecutive driving cycles in Fig.~\ref{fig:scaling}.
Since the system is initialized to the classical trajectory, $\braket{\hat{I}_{qp}}|_{t=0}$ corresponds to the value that the classical memristor returns after each driving cycle. Note that the memory-dependent damping in the input $\braket{\hat{V}}$ determines the decay and, hence, has the same scaling as the output. As briefly mentioned in the main text, the magnitude-scaled parametric hysteresis curves do not noticeably decay after 10 cycles for the orders of magnitude of the $T_1$-times between 1 ms and 10 $\mu$s. This is noticeable in Fig.~\ref{fig:scaling}, where the quasiparticle current after 10 cycles does not return to its original value when $\log [S_{qp}(\omega_{10})/\omega_{10}] > -2$.

\section{Effect of initialization on the hysteretic behavior and memory quantifier}

Here, we provide a qualitative interpretation for the varied hysteretic curves displayed in Fig.~3 of the main text. As in Fig.~2 of the main text, initially assuming a nonzero voltage over the SQUID and a negligible inductive phase difference yields hysteresis with $\pi$-rotational symmetry in the current-voltage plane (blue curve). On the other hand, assuming vanishing initial voltage and nonzero inductive phase difference destroys the hysteretic behavior (red curve). Finally, assuming that both are nonvanishing allows for voltage-asymmetric hysteresis (black curve). Each case can be seen as an example of the time-symmetry of the quasiparticle current between two consecutive crossings of the zero-energy point. With sinusoidal driving, both the voltage and the memristance are time symmetric, whereas the symmetry of their product, the quasiparticle current, depends on the specific initialization. Hence, tuning the initial conditions yields a variety of different memristic switchings which can, in general, be made voltage asymmetric. The parametric current-voltage characteristics can also be tuned by resorting to different bias protocols which may help to achieve desired behavior.

The memory quantifier for Fig.~2 in the main text decays pairwise linearly from $N_m^{1}/N_m^{cl} = 0.9960$ to $N_m^{19}/N_m^{cl} = 0.8511$, where $N_m^{cl}$ is the value for the classically initialized weak-damping solution, during the simulation. In the weak-damping regime, the decay is negligible. For the different initializations in Fig.~3 of the manuscript, the weak-damping solutions yield $N_m/N_m^{cl} = 1$ (blue), $N_m/N_m^{cl} \propto 10^{-7}$ (red), and $N_m/N_m^{cl} = 1.3409$ (black). Curiously the last case yields the same value for both asymmetric loops.

\section{Area of a hysteresis loop as a memory measurement} \label{area}

For a classical voltage-controlled  memristive system, the current response is generally given by~\cite{pieee97/1717}
\begin{equation}
I(t) = G(\gamma(t),V(t),t) V(t),
\label{eq:Icl}
\end{equation}
where $G(\gamma,V(t),t)$ is the memductance, which depends on the instantaneous value of the input voltage $V(t)$, the accumulated value of the memory variable $\gamma(t)$, and the time $t$ in a parametric manner. The memory variable update is defined by
\begin{equation}
\partial_t  \gamma(t) = f(\gamma(t),V(t),t),
\label{eq:fcl}
\end{equation}
where the update function $f$ depends on the past evolution, the input and any external parametric driving. Notice that the general definitions used above allow for tracking a specific state variable $\gamma$ rather than absorbing the total response into a generalized memory variable $\gamma_x$ such that $G(\gamma_x) = G(\gamma,V,t)$ where $\gamma_x = f_x(\gamma_x,V,t)$ now updates the new variable. This absorption can always be done, since memductance is a system property depending on the input and external driving via the state variable. 

Define $\mathcal{C}$ as any closed curve in the $(I,V)$-space, hence the area enclosed by the curve is determined by Green's theorem as
\begin{equation}
A = \frac{1}{2} \oint_{\mathcal{C}} (VdI-IdV) = \frac{1}{2} \oint_{\mathcal{C}} V^2 d \left( \frac{I}{V} \right)  = \frac{1}{2} \oint_{\mathcal{C}} V^2 d G,
\end{equation}
where the explicit dependences given by the response in Eq.~(\ref{eq:Icl}) are omitted for clarity. Transforming to temporal space, the integration takes the form
\begin{equation}
A = \frac{1}{2} \int_0^T dt V^2(\partial_{\gamma} G \partial_t \gamma + \partial_V G \partial_t V + \partial_t G),
\label{eq:A_cl_full}
\end{equation}
where we have fixed the loop to begin at time $t=0$ and end at $t=T$ when the initial point in the $(I,V)$-space is reached again. Assuming pinched hysteresis, the periodicity condition is conveniently written as $V(nT) = 0$, $n \in \mathbb{Z}$. Due to this condition, the second term in Eq.~(\ref{eq:A_cl_full}) yields
\begin{equation}
\frac{1}{2} \int_0^T dt V^2 \partial_V G \partial_t V = \frac{1}{2} \int_{V(0)}^{V(T)} dV V^2 \partial_V G = 0,
\end{equation}
implying that any instantaneous voltage dependence in the current response does not contribute to the area. By making use of Eq.~(\ref{eq:fcl}), the area is
\begin{equation}
A = \frac{1}{2} \int_0^T dt V(t)^2[\partial_{\gamma} G(\gamma,V,t) f(\gamma,V,t) + \partial_t G(\gamma,V,t)],
\label{eq:A_cl_final}
\end{equation}
where the first term in the integrand corresponds to the response related to the selected memory variable and the second term accounts for any remaining explicit time dependence of the conductance.

Using the notation above, any non-linear conductor $G=G(V)$ can be defined via $\gamma = V$ such that $f(\gamma) = \partial_t V$. Equation (\ref{eq:A_cl_final}) then yields
\begin{equation*}
A = \frac{1}{2} \int_0^T dt V(t)^2\partial_V G(V) \partial_t V = \frac{1}{2} \int_{V(0)}^{V(T)} dV V^2\partial_V G(V) ,
\end{equation*}
which is zero, $A=0$, due to the periodicity. In other words, purely non-linear conductance does not produce hysteresis, and cannot generate a non-zero loop area. The memory-variable related term in Eq.~(\ref{eq:A_cl_final}) can be rewritten as
\begin{widetext}
\begin{equation}
\begin{split}
\frac{1}{2} \int_0^T dt V(t)^2\partial_{\gamma} G(\gamma,V,t) f(\gamma,V,t) &= \frac{1}{2} \left[ \int_{0}^{T} V(t)^2 \partial_{\gamma} G(\gamma,V,t) \int_0^T dt f(\gamma,V,t) \right. \\ & \left. - \int_0^Tdt \int_0^t d\tau f(\gamma,V,\tau) \partial_t [V(t)^2 \partial_{\gamma} G(\gamma,V,t)] \right] \\ &= -\frac{1}{2} \int_0^T dt \Delta \gamma(t) \partial_t [V(t)^2 \partial_{\gamma} G(\gamma,V,t)],
\end{split}
\label{eq:A_second}
\end{equation}
\end{widetext}
where $\Delta \gamma(t) = \gamma(t) - \gamma(0)$ is the change in the memory variable from its initial value, we executed integration by parts after the first equality, and the first term after the first equality vanishes due to the periodicity condition. Applying the results in Eqs.~(\ref{eq:A_cl_final}) and (\ref{eq:A_second}) yields the definition of the loop area as a memory quantifier given in the main text for our superconducting circuit.

\end{document}